# Inorganic Surface Passivation of PbS Nanocrystals resulting in Strong Photoluminescent Emission.

M. J. Fernée[†], A. Watt, J. Warner, S. Cooper, N. Heckenberg, H. Rubinsztein-Dunlop.

*Center for Quantum Computer Technology, Department of Physics, University of Queensland, Brisbane, Queensland 4072.*

[†] *email: fernee@physics.uq.edu.au*

**Abstract**

Strong photoluminescent emission has been obtained from 3 nm PbS nanocrystals in aqueous colloidal solution, following treatment with CdS precursors. The observed emission can extend across the entire visible spectrum and usually includes a peak near 1.95 eV. We show that much of the visible emission results from absorption by higher-lying excited states above 3.0 eV with subsequent relaxation to and emission from states lying above the observed band-edge of the PbS nanocrystals. The fluorescent lifetimes for this emission are in the nanosecond regime, characteristic of exciton recombination.

**Introduction**

Colloidal nanocrystal (NC) quantum dots are nanometer sized semiconductor crystals in which at least one of the intrinsic charge carriers undergoes some degree of quantum confinement. In NCs this is usually seen as a structured absorption spectrum and a blue shift of the absorption band-edge with respect to the bulk semiconductor band-edge. The chemistry of NC synthesis has matured rapidly, with the ability to control the size, structure and shape of a wide variety of semiconductor materials using a diverse range of techniques [1-3]. Hitherto, one of the most useful properties of NCs is the ability to tune the absorption band-edge and consequently the fluorescent emission wavelength via precise control of the nanocrystal size, which controls the confinement energy of the charge carriers [4].

Due to the small size of NCs and hence their relatively large surface to volume ratio, the effect of the nanocrystal surface on the charge carriers cannot be neglected [5]. In practice the surface properties must be modified in some way in order to both control nanocrystalline growth [3,6,7] and reduce or prevent the charge carriers interacting with the surface, thus realizing effective quantum confinement. Surface passivation with various organic ligands [8-10] or epitaxial overcoating with a wide band-gap semiconductor [5,11-14] can be used to enable efficient radiative recombination of charge carriers. Photoluminescence quantum efficiencies approaching unity have been achieved [12,14-16], indicating that non-radiative recombination pathways can be effectively suppressed using these strategies.

Lead chalcogenide nanocrystals are currently generating a great deal of interest [10,15-18] as materials that are tunable throughout the near infrared region where they could be used for communications applications. Furthermore, the lead chalcogenides have exceptionally large exciton Bohr radii [17], which make them very susceptible to charge-carrier quantum confinement effects over a broad range of nanocrystal sizes. It has been suggested by Wise [17] that these nanocrystals are especially suitable for studying the strong quantum confinement regime due to their large excitonic Bohr radii and nearly identical charge carrier wavefunctions [19]. However, good surface passivation is necessary in order for lead chalcogenide nanocrystals to be made sufficiently robust and to prevent the intrinsic charge-carriers interacting with the nanocrystal surface.

Lead sulfide (PbS) in particular is a narrow band-gap semiconductor with highly symmetric energy level structure that can readily be prepared in the strong quantum confinement regime for both charge-carriers. For example, colloidal suspensions of PbS NC's can be prepared with a relatively narrow size distribution, exhibiting three pronounced "excitonic" peaks in the absorption spectrum [20-23] which are separated by more than 1 eV. However, it has been shown that the observation of distinct excitonic features in the absorption spectrum is strongly related to the surface properties, as attempts to use a range of organic ligands for surface passivation result in a relatively structureless absorption spectrum, even for extremely narrow size distributions [24]. This behaviour has been attributed to the incomplete passivation of the surface interactions with both charge carriers, resulting in one charge carrier strongly interacting with the surface, thus promoting charge separation [24]. The strong surface interactions in small PbS NC's also promote non-radiative recombination of charge carriers, resulting in notoriously weak photoluminescence. So far, the most successful technique for restricting surface interactions has been to grow larger NCs, which have a correspondingly smaller surface to volume ratio and a lesser degree of quantum confinement [25]. Otherwise, the weak photoluminescence of the smallest PbS NC's has restricted the study of strong quantum confinement to various nonlinear and absorption based techniques [22,24,26].

In this paper we show that it is possible to passivate the surfaces of PbS NC's in an aqueous colloidal solution by overcoating with cadmium sulfide (CdS), a higher band-gap semiconductor. We note that CdS has a different crystal structure than that of PbS, which generally precludes effective epitaxial overcoating. However, there is a partial crystalline compatibility as both materials can grow in cubic structures based on an fcc arrangement of atoms with similar lattice constants. This allows effective epitaxial growth on at least one crystal surface [27]. We believe that it is this partial compatibility which allows surface-passivation and as a result, significant photoluminescent emission can be achieved.

**Motivation**

Surface passivation by epitaxial overcoating a core NC with a large band-gap semiconductor has been developed for a number of NCs. For example, with CdSe NCs, epitaxial overcoating with either ZnS [11,12], ZnSe [14] or CdS [13] results in strong band-edge fluorescence with a pronounced improvement in the fluorescence quantum yield. In all cases, good epitaxial overcoating requires the use of a wide band-gap material with compatible lattice constants and crystal structure. This implicitly means using a material with the same crystal structure so that epitaxial overcoating can be accomplished in all crystal directions with the same efficiency.

The lead chalcogenides have a highly symmetrical rocksalt crystal structure, which is not common amongst most semiconductor materials. One possible group of compatible materials with both a wide band-gap and a rocksalt crystal structure are the alkaline-earth sulfides. However, solution phase syntheses of these materials has only recently been reported [28] and requires much further work. Alternatively, we noticed that another possibility might be realized with the zincblende structure of either ZnS or CdS. The zincblende structure, like the rocksalt structure, is also a cubic structure based on a fcc arrangement of constituent atoms and differs only in the coordination of the nearest-neighbour atoms. Rocksalt is a higher symmetry structure with six nearest-neighbour atoms, while the zincblende structure has only four. However, in both cases the (111) set of crystal planes is terminated by a single atomic species and is thus structurally identical. Therefore epitaxial growth on any (111)-terminated surface of a PbS NC is possible [27]. However, the compatible fcc arrangement of each atomic constituent suggests that to a first approximation, *any* surface that is completely terminated by a single atomic species should be indistinguishable for either rocksalt or zincblende structures. This naïve approximation however does not take into account the effect that bond reorientation may have on this compatibility.

PbS NCs can readily be made in the strong quantum confinement regime with an absorption band-edge near 600 nm, which is in a spectral region achievable with CdSe NCs. As both ZnS and CdS have been used for effective surface passivation of CdSe NCs, they may also provide an effective means for surface state passivation of PbS NCs, resulting in visible luminescent emission. We chose CdS (zincblende 5.82 Å) for potential epitaxial overgrowth due to the small lattice mismatch with PbS (rocksalt 5.91 Å). Some degree of fluorescence activation was also obtained with ZnS, but we found it far less effective than CdS. In figure 1 we show a radial potential model associated with a PbS core NC surrounded by a CdS shell, which is in turn surrounded by an extremely wide band-gap material such as water. The potentials were determined by combining the bulk band-gaps of the two semiconductors (at room temperature) and using the appropriate conduction band offset[*] [29]. According to this scheme the conduction band offset between PbS and CdS results in a shallower potential for the electron than for the hole, so the electron wavefunctions are not likely to be as well confined as the hole wavefunctions. For example, we indicate the estimated energies of the charge-carriers within the bare PbS NC corresponding to the first, and second excitonic absorption maxima (according to the isotropic-band model [19]) for a spherical 3 nm diameter PbS NC. While the lowest excited levels just fall within the CdS confinement potential, the higher excited levels are no longer well confined by the CdS overcoat. The electron wavefunction should be most strongly affected by the shallowness of the confinement potential, especially since the electron effective masses in both PbS and CdS are similar (0.16 $m_0$ and 0.165 $m_0$ respectively [30]). Therefore, the electron wavefunction is likely to significantly penetrate any CdS overcoat, suggesting that further passivation of the CdS surface may be necessary.

**Sample Preparation**

Surface-passivated PbS (SP-PbS) NCs are made according to the following procedure: Firstly aqueous colloidal suspensions of PbS NCs of approximately 3 nm diameter were prepared according to the method of Nenadovic and co-workers [21]. Surface-passivation of the PbS NCs was then accomplished by using a liquid phase epitaxial overcoating procedure that essentially adapts a well-known method for CdS NC synthesis as described by Spahnel and co-workers [31].

Typically, 100 ml of $2\times10^{-4}$ M solution of lead acetate (AR grade, Ajax) in 0.2% polyvinyl alcohol (22000 MW, BDH) was prepared using ultra-pure water (Millipore 18.2 MΩ cm). All chemicals were used directly without further purification. The solution was degassed by bubbling with Argon for 20 min, before the addition of 0.8 ml hydrogen sulfide gas. The amount of added $H_2S$ corresponds to more than a 2:1 molar excess and was used to promote the growth of PbS NCs with sulfur terminated surfaces. The solution was reacted for 5 minutes, forming a deep red PbS NC solution. The CdS overcoating procedure was then carried out as follows: While maintaining a pH of less than 5 in order to suppress individual CdS NC growth [32], 2 ml of a degassed 0.01 M sodium hexametaphosphate solution was added with rapid stirring. This was followed by the injection of a quantity of degassed $CdCl_2$ (AR grade, Ajax) solution sufficient to provide $Pb^{2+}$ to $Cd^{2+}$ ratio of 1:2. The injection of the $CdCl_2$ solution was either done slowly over a period of 2 minutes using a 0.01 M concentration solution or rapidly using a 0.1 M concentration. The solution was then reacted for 15 minutes with constant stirring. At this stage visible photoluminescence was detectable by eye when the solution was illuminated by a UV fluorescent lamp. A final photoluminescence activation process was usually performed [31] whereby the pH of the solution was raised above 10.5 by injecting 2 ml of a 0.1 M NaOH solution followed by the rapid injection of an excess 0.5ml of 0.1 M $CdCl_2$ solution. After this final stage a strong visible photoluminescence resulted. In general these SP-PbS solutions maintain their luminescent properties for many months if kept in the dark. However, the absorption pales significantly over one or two weeks before stabilizing.

---

[*] The relative band offsets for PbS and CdS are not well known. We have chosen the offset presented by Vogel *et al.* [29] in their PbS sensitized solar-cell work. Alternative studies of the PbS/CdS system include the bulk heterojunction work of Watanabe and Mita [27] and the superlattice work of Musikhin *et al.* (Superlatt. Microst. **15**(4), 495, (1994)). While the results of these studies vary considerably, they are all consistent in having a small conduction band offset and a large valence band offset.

The two variations in the addition of the $CdCl_2$ solution produce very different photoluminescent behaviors that are described in more detail below. Further minor variations to the above procedure, including varying the $Pb^{2+}$ to $Cd^{2+}$ ratio, also produce SP-PbS NCs displaying qualitatively similar luminescent properties.

**Characterization Procedures**

Bulk absorption studies were conducted using a UV-Visible Spectrophotometer (Perkin-Elmer Lambda 40). Measurements were taken using a 10mm cuvette (both UV quartz and PMMA) with the samples diluted with ultra-pure water in order to ensure that absorption is kept within the Beer-Lambert limit. Appropriate solutions of PVA, sodium hexametaphosphate and PVA/hexametaphosphate mixtures were used for background subtraction in order to produce the final absorption profiles.

Some of the photoluminescence (PL) data (figure 3 and figure 4) were obtained using a commercial fluorimeter (Perkin-Elmer LS50B) with the slits set for a 10 nm bandpass in both the excitation and detection channels and without using any correction for Raman and Rayleigh scattering. PL and photoluminescence excitation (PLE) spectroscopy for figures 5 and 6 were conducted using a commercial fluorimeter (Spex Fluoromax3). Here the spectra obtained were corrected for Raman and Rayleigh scattering where possible using a non-fluorescing scattering solution to provide the appropriate correction files. The slits for the excitation and detection channels were set to obtain a 3 nm bandpass. Time-resolved PL decays were obtained using a time-correlated single photon counting spectrometer (Picoquant Fluotime 200) with femtosecond pulses sourced from a frequency-doubled modelocked Ti:sapphire laser (Spectra Physics Tsunami). The bandwidth of the detection path was set at 16 nm by using 2 mm slits with a dual grating subtractive monochromator.

**Spectroscopic Investigation**

The effect of the addition of CdS precursors into a colloidal suspension of PbS NCs is monitored using absorption in figure 2. In general, as the PbS NCs are made without any pH buffering, the pH of the resulting colloidal suspension has been found to lie between 3 and 4. It is known that at acid pH, the introduction of CdS precursors into another NC solution favors increased growth of the existing NCs rather than new CdS NC growth [32]. In figure 2 (a) we show the absorption spectra for pure PbS NCs, SP-PbS NCs (made with a Pb:Cd ratio of 1:2) and pure CdS NC's. The second derivative of each curve is shown in figure 2 (b) in order to amplify any structure that isn't readily detectable in the direct absorption spectra. The PbS based NC solutions have the familiar three peaked structure attributed to exciton absorption resonances in PbS NCs [20-23]. The SP-PbS spectrum shows a slight blue-shift of the peaks and an increased high energy absorbance, but is otherwise directly related to the spectrum obtained from pure PbS NCs. This similarity is also borne out in the second derivative curves. We also show the absorption spectrum of the colloidal CdS NC solution that results when there are no PbS NCs present (ie. in an aqueous 0.2% PVA medium without using any pH buffering [31]), in order to provide a CdS NC absorption signature. Using this absorption spectrum for comparison, we see no evidence of coincidental CdS NC formation in the SP-PbS NC absorption and second derivative spectra. This provides a strong indication that individual CdS NC growth is indeed suppressed during the process of surface-passivation.

We can clearly see from both the absorption and second derivative spectra that the process of surface passivation results in a blue-shift of the band-edge (eg. figure 2(a) inset) and peaks of the PbS NC spectrum. The presence of a semiconductor shell as grown by liquid phase epitaxy usually results in a red-shift of the absorption spectrum [11-14], while a blue-shift such as we observe can indicate the presence of alloying at a core/shell boundary [5,12]. This is not so surprising given the difference in crystal structures already mentioned above. While PbS and CdS are stable phases at room temperature and do not alloy [33,34], they may possess a highly disordered boundary phase [35]. However, in our case the origin is most likely to result from the displacement of $Pb^{2+}$ ions at the NC surface during the addition of $Cd^{2+}$ ions at acid pH, causing partial erosion of the NC surface [36]. We observe this erosion as a gradual increase in the $Pb^{2+}$ ion concentration and concomitant degradation of the PbS NC absorption spectrum. Raising the pH of the solution into the basic regime (typically pH 10.5 to 11) immediately after the $Cd^{2+}$ and $S^{2-}$ ions have reacted (15 minutes typically) seems to prevent further erosion.

The process of surface passivation with CdS can result in a wide variety of different photoluminescent (PL) spectra ranging from a deep red emission to a broad white-green emission. These two extremes are represented in figures 3 (a) and (b), where we also include a typical SP-PbS NC absorption spectrum for comparison. We see in figure 3 (a) an abrupt onset of the deep red PL at the band-edge of the SP-PbS NCs and so refer to this type of emisson as band-edge emission. In contrast, the onset of emission for the white-green emission occurs above the SP-PbS absorption band-edge and so we refer to this type of emission as above-band-edge. Two luminescent SP-PbS solutions are shown in figure 3 (c) as an example of the variation in emission that can be obtained.

The surface-passivation technique that we describe is a two-part procedure where first $S^{2-}$ and then $Cd^{2+}$ ions are introduced at acid pH in an attempt to passivate the dangling surface bonds of the PbS NC, then the pH is raised to a strongly alkaline condition with the injection of NaOH followed by a large excess of $Cd^{2+}$ ions in order to form a secondary passivating $Cd(OH)_2$ layer. We refer to the final process of $Cd(OH)_2$ growth as the activation phase, while prior to this the SP-PbS NCs are described as unactivated. We note that we are unable to detect PL from bare PbS NCs or CdS NCs in aqueous colloidal solution at acid pH (ie. our unactivated condition). Therefore, we can attribute the unactivated PL to PbS NCs that have been surface passivated with CdS. In general we observe reasonably strong PL emission from unactivated SP-PbS NCs, which is subsequently enhanced by a factor of between 2 and 4 upon activation. The effect of this process of activation is illustrated in figure 4 (a) using a SP-PbS NC solution with a Pb:Cd ratio of 1:2 used in the initial unactivated phase. The two PL were pumped at 3.18 eV and have been scaled so that the low energy tails are coincident. It is immediately obvious that the activation procedure acts to increase the above-band-edge portion of the PL spectrum. Further insight into this process is gained by plotting the ratio of the activated to the unactivated PL spectrum in figure 4 (b). Here we find that the activation procedure does indeed favor an increase in the high-energy portion of the PL spectrum, but we also find evidence for a sharp onset of PL emission near 2.9 eV. This sharp onset is not evident from the individual PL spectra alone.

For CdS NCs, overcoating with a $Cd(OH)_2$ layer is known to passivate surface electron traps [13,31]. This is also the likely mode of operation here and suggests that the initial surface-passivation procedure with CdS primarily passivates hole traps on the PbS NC surface, while only partially passivating the electron traps. This is consistent with the confinement potentials plotted in figure 1, which show poorer confinement of the electron wavefunctions due to the conduction band offset for a PbS/CdS heterojunction.

Photoluminescence excitation (PLE) scans from a PbS/CdS NC solution made with a Pb:Cd ratio of 1:2 are shown in figure 5. A photoluminescence (PL) spectrum obtained from a different but representative 1:2 solution pumped at 3.18 eV is included here merely as a guide. In the PLE curve monitoring the band-edge emission at 1.9 eV we see a three peaked structure (peaks at approximately 2.2 eV, 3.4 eV and 4.5 eV) which closely resembles the three peaks observed in the SP-PbS NC absorption spectrum. Similarly, the PLE spectrum monitoring the above-band-edge emission at 2.43 eV also shows what are clearly two of the three peaks, although slightly blue-shifted. The 2.43 eV PLE spectrum is representative of that obtained at all points along the above-band-edge emission spectrum. Therefore we find that the above-band-edge portion of the PL spectrum does not represent emission from states that can be directly excited, but rather emission that is somehow coupled to, and significantly red-shifted from, a series of higher energy states. In general the similarity between the PbS NC absorption spectrum and the PLE spectra show that the absorption of radiation and subsequent emission is strongly influenced by PbS NCs.

Further insight into the origin and nature of the above-band-edge PL is obtained from the time-resolved PL decay curves that are presented in figure 6 (a). This data was obtained from SP-PbS NCs embedded in a thin PVA film, which was mounted in a liquid nitrogen cryostat. Open circles shown on the PL spectrum in figure 6 (b) mark the energies at which the PL was monitored for the time-resolved PL study. The use of a logarithmic count axis clearly shows that these fluorescence decay curves are highly multi-exponential. However a significant component of the PL emission occurs with lifetimes in the 1-3 ns range, which is characteristic of excitonic processes [37]. There is also a continual increase in the proportion of components with longer lifetimes as the PbS band-edge is approached, which is probably due to an increasing number of trap states of various types at the PbS/CdS interface. We note that the effect of the different supporting matrix and lower temperature serves to significantly increase the photoluminescence yield while decreasing the contributions from long-lived processes and increasing the lifetimes of the nanosecond components. The combined effect thus enhances the nanosecond processes at the expense of the longer-lived processes. Nevertheless the nanosecond components are still observed at room temperature in an aqueous solution.

**Discussion**

The surface passivation procedure that we describe is relatively robust, resulting in strongly luminescent solutions. However, we have not yet determined a reliable procedure under which the luminescence can be tuned from strong band-edge to strong above-band-edge. Our experiments point to a number of linked parameters that affect the surface-passivation process, such as solution pH, temperature, rate of $Cd^{2+}$ ion addition, $Cd^{2+}$ ion and polyphosphate concentrations and the initial $S^{2-}$ ion concentration. However solutions that produce a deep red band-edge emission are generally made with a slow addition of $Cd^{2+}$ ions over many minutes and using a $10^{-2}$ M solution, while the strong above-band-edge emission requires a rapid addition of $Cd^{2+}$ ions usually using a $10^{-1}$ M solution.

Of the two types of emission that we have categorized, the band-edge emission is most likely to contain a significant proportion of direct exciton recombination at the SP-PbS NC band-edge. This is suggested primarily by the steep onset of emission at the band-edge region, as shown in figure 3(a). We have also found that the edge of the emission corresponds to a Gaussian FWHM of 250 meV, which matches closely the Gaussian FWHM of the absorption band-edge. However there exists a large proportion of this emission that extends into the near infrared region, which is likely due to one or more trapped charge carriers. Time-resolved PL measurements in this region exhibit decay times in the microsecond regime, characteristic of trapped charge-carrier recombination. Nevertheless, it is possible that direct exciton recombination at the band-edge may also exhibit microsecond lifetimes similar to what has been found for PbSe NCs [15,16].

The above-band-edge emission is not as straightforward to explain. PLE spectra clearly link this emission to absorption by higher excited states within SP-PbS. However, we have shown that the luminescence onset is as abrupt as the absorption onset, while the peak emission is strongly red-shifted from the absorption edge. Such behavior is often associated with luminescent impurity states [38,39] and trap states. However, such processes usually have long luminescence lifetimes. On the contrary, we find that this emission is dominated by short nanosecond lifetimes, more characteristic of excitonic emission. We suggest that this emission is closely linked to charge separation in a core/shell NC model [40,41]. In addition, partial crystal structure incompatibilities are likely to result in interface and defect states, which would be expected to influence the luminescence behavior of this system.

As our current studies of the SP-PbS NCs have been primarily limited to spectroscopic investigation, we are only able to infer structural information. While we have referred to a procedure for surface-passivation, we believe that our procedure actually produces a core/shell structure. In fact most of our data is consistent with such a structure, which we summarize as follows: The absorption data shows that the addition of a two-fold molar excess of $Cd^{2+}$ and $S^{2-}$ ions produced no indication of CdS NC formation, but retained the essential PbS NC absorption structure. The observation of PL from an unactivated solution, shows that surface-passivation is accomplished by the addition of CdS precursors only, which suggests the formation of a CdS shell. The increase of the PL emission following the activation process is consistent with the passivation of electron traps on a CdS surface [13,31]. The similarity of the PLE spectra with the absorption spectrum suggests that absorption by the luminescent species is strongly influenced by a PbS NC core. Finally, we have made both pure CdS NCs in low pH conditions and mixed PbCdS NCs and find no emission before activation and only weak PL emission when activated. This confirms that neither CdS nor Pb contaminated CdS NCs are responsible for the observed PL when unactivated or the strong visible emission when activated with $Cd(OH)_2$.

The PbS NC synthesis we have used currently restricts our spectroscopic investigation of this system as the synthesis has no potential for varying the size of the PbS NCs produced. Therefore, we are unable to show any size-dependence in the luminescence behavior. Quantum confined excitonic effects are most satisfactorily demonstrated through size dependent spectra. Alternatively we have provided less conventional evidence, including a study of the effect of activation with $Cd(OH)_2$ and luminescence lifetime data.

**Conclusion**

We have shown that it is possible to use epitaxial overgrowth techniques in order to passivate the surface states of PbS NC's. Due to the difference in crystal structure between PbS and CdS, it is not obvious that epitaxial overcoating will occur, nevertheless highly luminescent SP-PbS NCs result. Apart from the expected band-edge emission we also find that these SP-PbS NCs can exhibit extremely broad visible emission that looks nearly white to the eye. We have established that most of this white emission appears to come from above the SP-PbS NC band-edge and results from absorption by higher excited excitonic states. Furthermore, the above-band-edge emission has a significant nanosecond component and so appears to be predominantly excitonic, rather than the result of one or more trapped charge carriers.

The ability to use a crystalline inorganic material for surface-passivation of a core material which has a different crystal structure, has important implications for the rapidly developing field of lead chalcogenide nanocrystal development in particular, and nanotechnology in general. While we have demonstrated the possibility of using such an approach, our results are only intended to show a proof of principle and therefore leave considerable room for optimization.

**Acknowledgements:**
This research was supported by the Australian Research Council (ARC).

**References:**

**[1]** See for example; Gaponenko S. V., "Optical Properties of Semiconductor Nanocrystals", Cambridge University Press, 1998.

**[2]** Yoffe A. D., Adv. In Physics **50**(1), 1 (2001).

**[3]** Puntes V.F., Krishnan K. M., Alivisatos A. P., Science **291**, 2115 (2001); Manna L., Scher E. C., Alivisatos A. P., J. Am. Chem. Soc. **122**, 12700 (2000) .

**[4]** Alivisatos A. P., J. Phys. Chem. **100**, 13226 (1996).

**[5]** Cao Y., Banin U., Angew. Chem. Int. Ed. **38**(24), 3692 (1999).

**[6]** Lee S.-M., Jun Y., Cho S.-N., Cheon J., J. Am. Chem. Soc. **124**, 11244 (2002).

**[7]** Patel A. A., *et al.*, J. Phys. Chem. B **104**, 11598 (2000).

**[8]** Kuno M., *et.al.*, J. Chem. Phys. **106**(23), 9869 (1997).

**[9]** Kapitonov A. M., *et al.*, J. Phys. Chem. B **103**, 10109 (1999); Kuno M., *et al.*, J. Chem Phys. **106**(23), 9869 (1997).

**[10]** Murray C. B., *et al.*, IBM J. Res. & Dev. **45**(1), 47 (2001).

**[11]** Hines M. A., Guyot-Sionnest P., J. Phys. Chem. **100**, 468 (1996).

**[12]** Dabbousi B. O., *et al.*, J. Phys. Chem. B **101**, 9463 (1997).

**[13]** Peng X., Schlamp M. C., Kadavanich A. V., Alivisatos A. P., J. Am. Chem. Soc. **119**, 7019 (1997).

**[14]** Reiss P., Bleuse J., Pron A., Nano Lett. **2**(7), 781 (2002).

**[15]** Wehrenberg B. L., Wang C., Guyot-Sionnest P., J. Phys. Chem B **106**, 10634 (2002).

**[16]** Du H., et al., Nano Lett. **2**(11), 1321 (2002).

**[17]** Wise F. W., Acc. Chem. Res. **33**, 773 (2000).

**[18]** Sashchiuk A., Langof L., Chaim R., Lifshitz E., J. Cryst. Growth **240**, 431 (2002).

**[19]** Kang I., Wise F. W., J. Opt. Soc. Am B **14**(7), 1632 (1997).

**[20]** Gallardo S., Gutierrez M., Henglein A., Janata E., Ber. Bun. Phys. Chem. **93**, 1080 (1989).

**[21]** Nenadovic M. T., Comor M. I., Vasic V., Micic O. I., J. Phys. Chem. **94**, 6390 (1990).

**[22]** Machol J. L., Wise F. W., Patel R. C. Tanner D. B., Phys. Rev. B **48**(4), 2819 (1993).

**[23]** Lu S., *et al.*, J. Mat. Sci. Lett. **17**, 2071 (1998).

**[24]** Guo L., *et al.*, Opt. Mat. **14**, 247 (2000).; Ai X. *et al.*, Mat. Lett. **38**, 131 (1999).

**[25]** Wundke K., *et al.*, Appl. Phys. Lett. **75**(20), 3060 (1999).

**[26]** Olkhovets A., Hsu R. C., Lipovskii A., Wise F. W., Phys. Rev. Lett. **81**(16), 3539 (1998); Krauss T.D., Wise F. W., Tanner D. B., Phys. Rev. Lett. **76**, 1376 (1996); Krauss T.D., Wise F. W., Phys. Rev. Lett. **79**, 5102 (1997); Krauss T.D., Wise F. W., Phys. Rev. B **55**, 9860 (1997).

**[27]** Watanabe S., Mita Y., Solid-State Elect. **15**, 5 (1972); Watanabe S., Mita Y., J. Electrochem. Soc. **116**, 989 (1969).

**[28]** Wang C., *et al.*, Chem. Phys. Lett. **351**, 385 (2002).

**[29]** Vogel R., Hoyer P., Weller H., J. Phys. Chem. **98**, 3183 (1994).

**[30]** CRC Handbook of Chemistry and Physics, Vol 83, (CRC Press, 2002).

**[31]** Spahnel L., Haase M., Weller H., Henglein A., J. Am. Chem. Soc. **109**, 5649 (1987); Eychmuller A., Hasselbarth A., Katsikas L. Weller H., J. Lumin. **48/49**, 745 (1991).

**[32]** Mews A., Kadavanich A. V., Banin U., Alivisatos A. P., Phys. Rev. B **53**(20), R13242 (1996).

**[33]** Holloway H., Jesion G., Phys. Rev. B 26(10), 5617 (1982).

**[34]** Boudjoul P., *et al.*, Chem. Mater. **10**, 2358 (1998).

**[35]** Zhou H. S., Honma I., Komiyama H., Haus J. W., J Phys. Chem **97**, 895 (1993).

**[36]** Moriguchi I., *et al.*, J. Chem. Soc., Faraday Trans. 94(15), 2199 (1998).

**[37]** Nirmal M., Norris D. J., Kuno M., Bawendi M. G., Phys. Rev. Lett. 75(20), 3728 (1995).

**[38]** Yang P., Lü M., Xü D., Yuan D., Zhou G., Chem. Phys. Lett. **336**, 76 (2001).

**[39]** Babin V., Krasnikov A., Nikl M., Stolovits A., Zazubovich S., Phys. Stat. Sol. (b) **229**(3), 1295 (2002).

**[40]** Jaskolski W., Bryant G. W., Phys. Rev. B **57**(8), R4237 (1998).

**[41]** Little R. B., *et al.*, J Phys. Chem. A **102**(33), 6581 (1998).

**Figure Captions:**

Figure 1. A potential energy diagram as a function of radius, r, representing both the conduction and valence band confinement potentials for a PbS/CdS core/shell NC structure. For comparison, the approximate electron and hole energy levels for a 3 nm PbS NC are indicated with dashed lines and the allowed optical transitions, corresponding to the first two exciton peaks, are indicated with adjoining arrows. These properties were obtained from the symmetric band model of Kang and Wise [26] and the dominant charge-carrier wavefunction symmetry is indicated by the atomic angular momentum nomenclature.

Figure 2. Absorption characterization (a) A comparison of absorption spectra for PbS NCs (black), SP-PbS NCs (blue) and CdS NC (red) are shown. (inset: expanded view of the absorption band-edge region indicated by a dotted box) (b) The second derivative of PbS NCs (black), SP-PbS NCs (blue) and CdS NC (red) absorption curves.

Figure 3. PL examples (a) Red band-edge PL spectrum (red) obtained from a SP-PbS solution made with an initial $Cd^{2+}$ to $Pb^{2+}$ ratio of 2:1 and slow injection of 0.01 M $CdCl_2$ over 2 minutes. The position of the emission is highlighted by a typical SP-PbS NC absorption spectrum (black). (b) Broad near-white PL spectrum (orange) obtained from a SP-PbS solution made with an initial $Cd^{2+}$ to $Pb^{2+}$ ratio of 2:1 and rapid injection of 0.1 M $CdCl_2$. The above-band-edge portion of this emission is indicated with reference to a typical SP-PbS NC absorption spectrum (black). (c) A photograph of the emission obtained using a hand-held UV-fluorescent lamp. Two different SP-PbS solutions are shown for comparison; a red band-edge solution and a yellow above-band-edge solution.

Figure 4. The effect of activation (a) PL spectra of both unactivated (black) and activated (blue) SP-PbS NC solutions ($Cd^{2+}$:$Pb^{2+}$ ratio of 2:1 and rapid injection of 0.1 M $CdCl_2$), pumped at 390 nm. The unactivated spectrum has been scaled so that the low energy tail is coincident with that of the activated spectrum in order to enhance the comparison. (b) The ratio of the activated spectrum divided by the unactivated spectrum, indicating the effect of the activation process is to enhance the high-energy part of the spectrum. However a sharp cut-off of the PL emission is also revealed.

Figure 5. PLE spectra obtained from a SP-PbS NC solution made with a $Cd^{2+}$ to $Pb^{2+}$ ratio of 2:1 and rapid injection of 0.1 M $CdCl_2$. The two curves represent monitoring the band-edge emission at 1.94 eV(black) and the above-band-edge emission at 2.43 eV (blue). A representative PL spectrum obtained from a similar SP-PbS NC solution is included for comparison (orange) with arrows used to indicate the emission energies used for the two PLE spectra.

Figure 6. Transient luminescent response (a) PL decays corresponding to emission from different points along the above-band-edge PL emission region. SP-PbS NCs (with a $Cd^{2+}$ to $Pb^{2+}$ ratio of 2:1 and rapid injection of 0.1 M $CdCl_2$) embedded in a PVA film at 77 Kelvin were used to obtain these data. The excitation was accomplished using 70 fs pulses centered at 380 nm obtained from a frequency doubled modelocked Ti:Sapphire laser. (b) The PL spectrum obtained from the same film at 77 K with pumping at 3.1 eV. Open circles on the curve indicate the emission energies used for monitoring the PL decays.

**Figure 1**

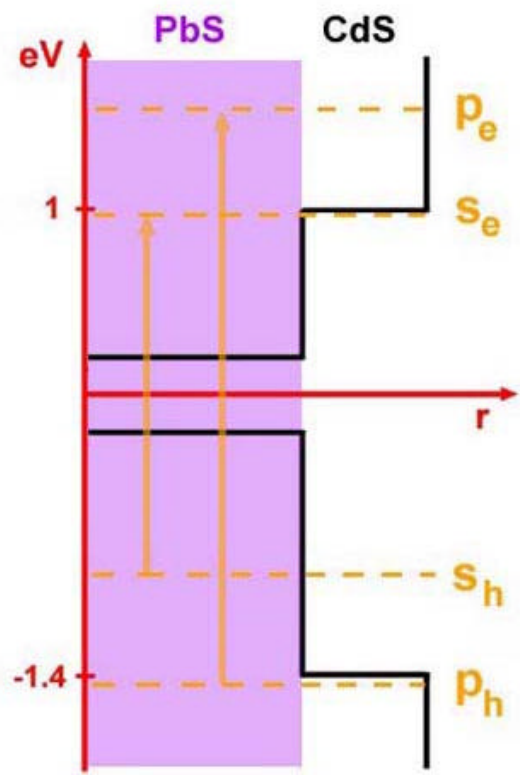

# Figure 2

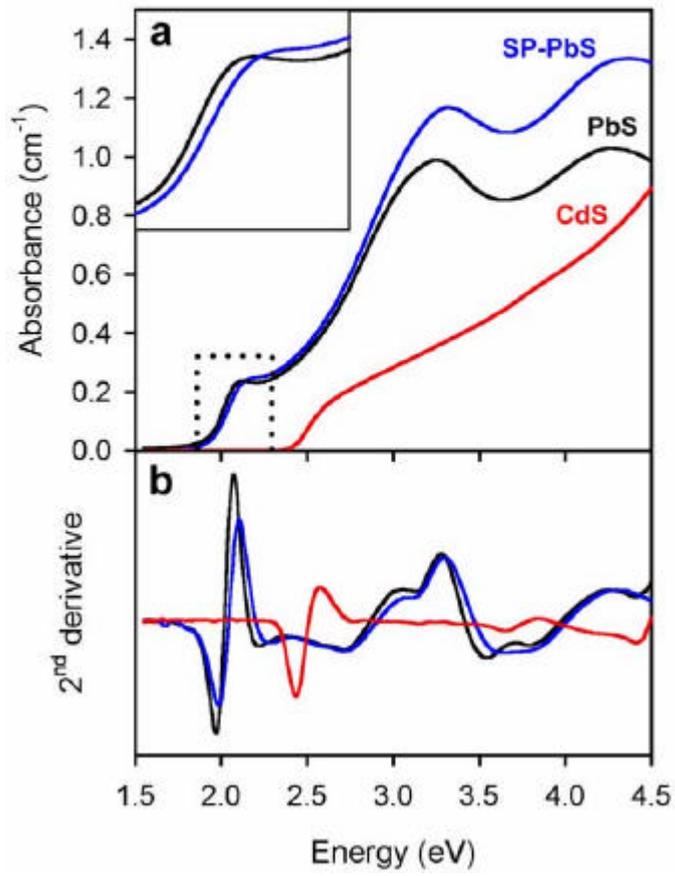

Figure 3

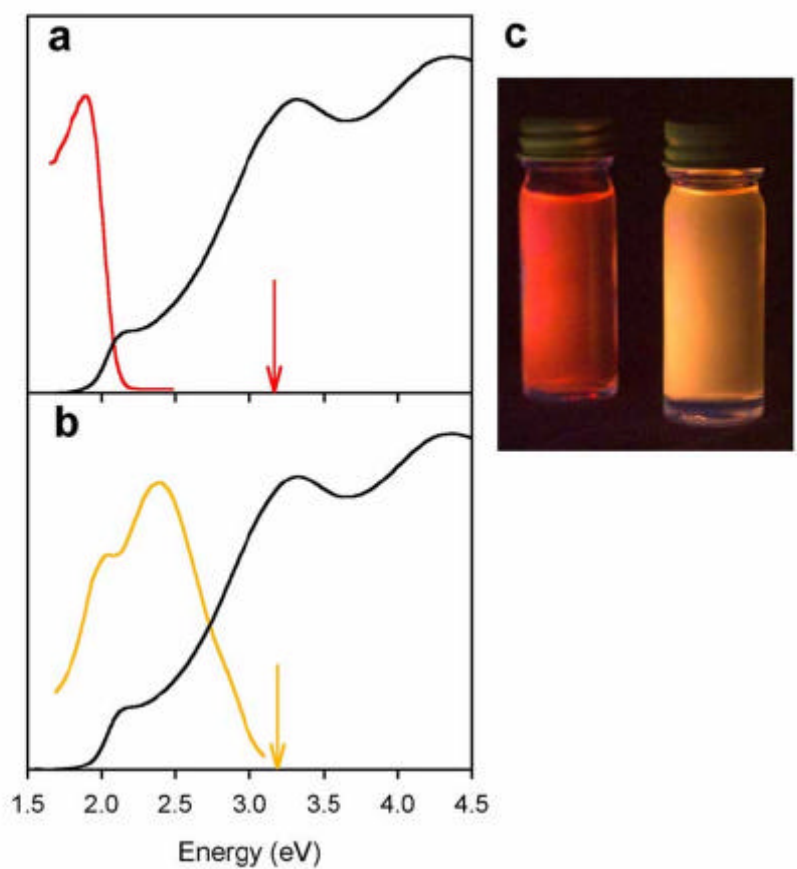

**Figure 4**

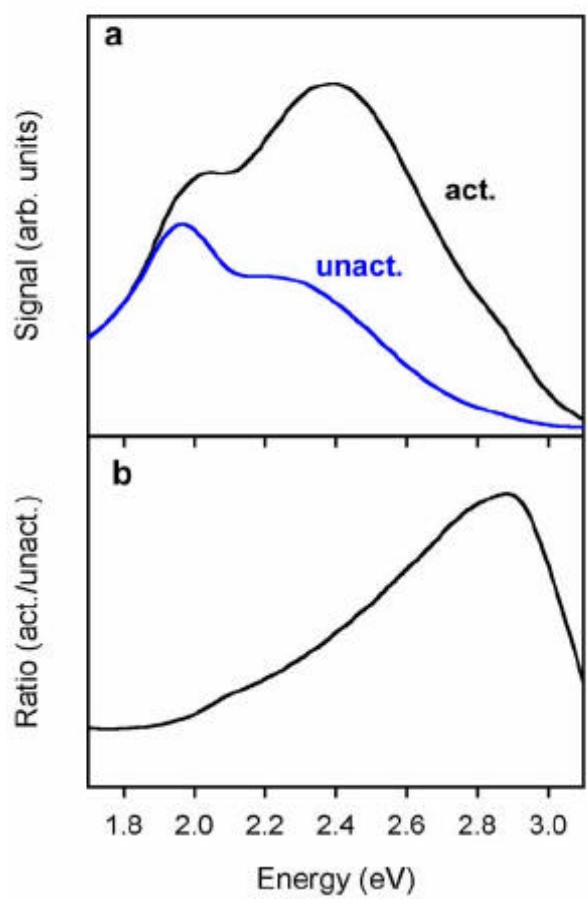

**Figure 5**

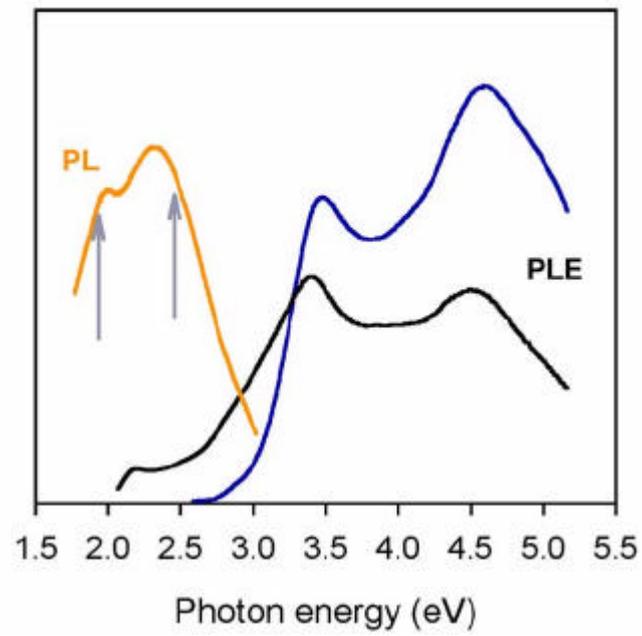

# Figure 6

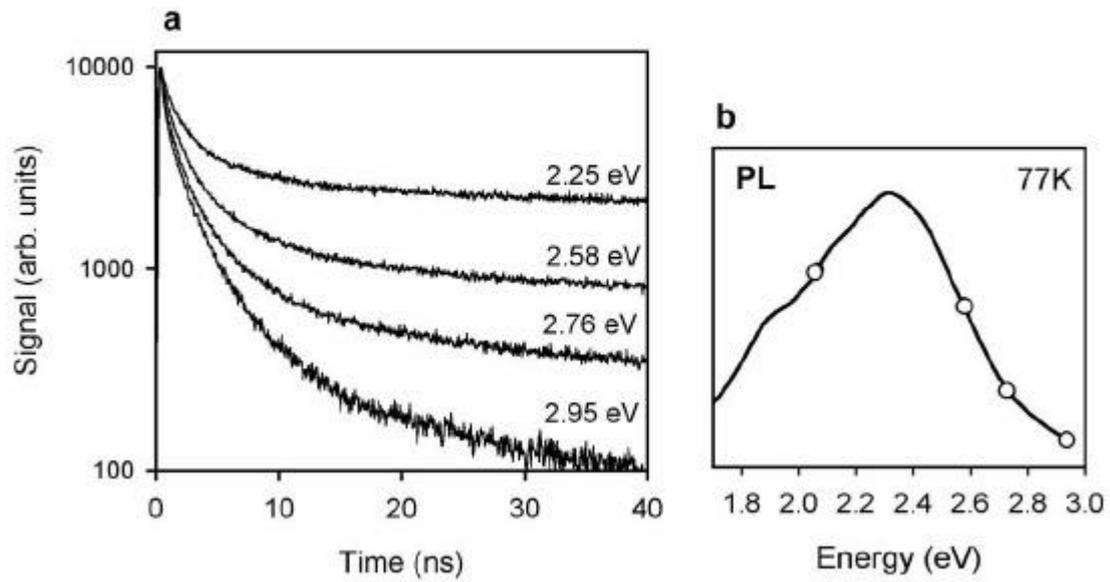